\documentstyle[12pt,epsfig]{article}
\begin{document}
\thispagestyle{empty} \vspace*{2cm}
\begin{center}
{\bf Spin Dynamics of Layered Triangular Antiferromagnets with Uniaxial Anisotropy}\\[1.cm]
\end{center}
\begin{center}
{Ranjan Chaudhury}\\
{S.N.Bose National Centre For Basic Sciences, Calcutta, India}\\
 ranjan@boson.bose.res.in
\end{center}
\vspace*{2cm}

 PACS:- 75.10.HK + 75.50.-y
\vspace*{0.3cm}

The spin dynamics of the semiclassical Heisenberg model with uniaxial anisotropy, on the layered triangular lattice with antiferromagnetic coupling for both intralayer nearest neighbor interaction and interlayer interaction is studied both in the ordered phase and in the paramagnetic phase, using the Monte Carlo-molecular dynamics technique. The important quantities calculated are the full dynamic structure function $S(\bf q, \omega)$, the chiral dynamic structure function $S_{chi}(\omega)$, the static order parameter and some thermodynamic quantities. Our results show the existence of propagating modes corresponding to both $S(\bf q, \omega)$ and $S_{chi}(\omega)$ in the ordered phase, supporting the recent conjectures. Our results for the static properties show the magnetic ordering in each layer to be of coplanar $3$-sublattice type deviating from $120$ degree structure. In the presence of magnetic trimerization however, we find the $3$-sublattice structure to be weakened alongwith the tendency towards non-coplanarity of the spins.

\section{Introduction}
  The layered triangular antiferromagnets with crystal field induced single ion uniaxial anisotropy are very important both from the viewpoint of behavior of magnetic materials as well as from the interest in spin models and universality classes in statistical mechanics$^{1,2}$. The materials $CsMnBr_2$, $VCl_2$, $VBr_2$ are some of the materials belonging to this class. They exhibit a complex competition involving geometric frustration and spin component degeneracy along the easy axis. The neutron scattering experiments on these materials both in the ordered phase as well as in the disordered phase by the experimentalists$^{3}$ provide a new impetus for theoretical study. The physical quantities of principal interest are the nature of static magnetic ordering and spin dynamics. In particular there are very interesting questions regarding the coplanarity and the sublattice structure of the magnetic ordering$^{3,4}$. In dynamics there are speculations about the chiral mode$^{4}$. The only reliable way at present to calculate the static correlations of the Heisenberg model is by the technique of Monte Carlo (MC) simulation. Furthermore the spin dynamics of this model can also be accurately studied only by using the technique of MC combined with Molecular Dynamics (MD). We have been motivated by the recent experiments to compute dynamic spin chirality correlator structure factor$S_{chi}(\omega)$ and total dynamic structure factor $S(q,\omega)$ for this model appropriate to layered triangular antiferromagnets both below and above Neel temperature. Besides, we also analyse the detailed nature of the magnetization vector(the conventional order parameter) and the effect of magnetic trimerization on the ordering pattern. Here we carry out our MCMD calculation with model parameters appropriate to Mn based systems with spin value(S) equal to $2$. We apply quantum correction to our semiclassical MCMD calculation to make our theoretical results comparable with the experimental ones.

  The plan of the paper is as follows:- In Sec.II we describe the model and the quantities of interest which we calculate; in Sec. III we very briefly describe the MCMD technique and our calculational procedure; in Section IV we display all the important results from our calculation; and finally in Sec. V we discuss some implications of our results.

\section{Model and Mathematical formulation}

  The model appropriate for our investigation is 

\begin{equation}
{\mathcal{H}}=\sum_{{\langle}ij{\rangle},{\langle}{\langle}ij{\rangle}{\rangle}} J_{ij}{\bf S}_i{\cdot}{\bf S}_j + D\sum_{i} {S_i^z}^2
\end{equation}

where $J_{ij}$ is positive and takes the value $J_1$ for $ij$ being nearest neighbors (i.e. for intralayer coupling) and the value $J_2$ for $ij$ being next nearest neighbors (i.e. interlayer coupling); $D$ is -ve.
  The definitions of most of the quantities of interest viz. $S(q,\omega)$, ${\bf M}$ etc. are very common and can be found in Ref.5. The new and very interesting quantity studied here viz. $S_{chi}(\omega)$ is defined as :-
\begin{equation}
S_{chi}(\omega)=\int\limits_{-\infty}^{\infty}dtC_{chi}(t)e^{-i{\omega}t}
\end{equation}
where $C_{chi}(t)$ is the scalar spin chirality correlator. It is defined as :-
\begin{equation}
C_{chi}(t)={\langle}{\chi}(0){\chi}(t){\rangle},\;
{\chi}(t)=\sum_{{\langle}ijk{\rangle}}{{\chi}^{p}}_{ijk}(t),\;
{\chi}(0)=\sum_{{\langle}lmn{\rangle}}{{\chi}^{p}}_{lmn}(0)
\end{equation}
where ${\langle}ijk{\rangle}$ and ${\langle}lmn{\rangle}$ represent the plaquettes in the triangular lattice. The total lattice scalar spin chirality ${\chi}(t)$ has been defined in terms of the local scalar spin chirality ${\chi}^{p}_{ijk}(t)$. The ${\langle}$  ${\rangle}$ occurring in the equation for $C_{chi}(t)$ have been used to represent the thermodynamic average. The sum in the definition of lattice chirality is taken over a plaquette in the anti-clockwise direction. The local spin chirality itself is defined as:-
\begin{equation}
{{{\chi}^{p}}_{ijk}}={\langle}{\bf S}{_i}{\cdot}{\bf S}_{j}X{\bf S}_k{\rangle}
\end{equation}
The other static quantity introduced in this paper is inverse degree of coplanarity $\gamma$ defined as:-
\begin{equation}
\gamma=[{\bf M}_{A}{\cdot}{\bf M}_{B}X{\bf M}_C]/[|{\bf M}_{A}||{\bf M}_{B}||{\bf M}_{C}|]
\end{equation}
where ${\bf M}_A$,${\bf M}_B$ and ${\bf M}_C$ are the magnetization vector order parameters in the three sublattices A,B and C respectively. We evaluate $\gamma$ for our model both in the absence and in the presence of trimerization.
                                                                             
\section{The MCMD Approach and Calculations}

  The basic equation governing the spin dynamics is the following:-
\begin{equation}
 \frac{d{\bf S}_{i}^{op}}{dt}={\frac{2\pi}{ih}}[{{{\bf S}_{i}}^{op}},{\mathcal{H}}]
\end{equation}

where the square bracket denotes the commutator. This equation leads to more simplified equations of motion :-
\begin{equation}
\frac{d{{\bf S}_{i}}}{dt}=\frac{2\pi}{h}{\sum_{j}}{J_{ij}}({{{\bf S}_{i}}}X{{{\bf S}_{j}}})+D{\frac{2\pi}{h}}[-{\bf x}{S^{y}}_{i}+{\bf y}{S^{x}}_{i}]
\end{equation}
 The more details regarding these can be found in Ref.s $5$ and $6$. We  take the semiclassical version of the above equation for spin dynamics$^{5}$. We then apply quantum correction to evaluated semiclassical $S_{chi}(\omega)$ and $S(q,\omega)$ following Windsor's prescription$^{5}$. The corrected $S_{chi}(\omega)$ takes the following form in this prescription:-
\begin{equation}
 S_{chi}^{QM}(\omega)=[\frac{2}{1+e^{-\frac{h{\omega}{\beta}}{2\pi}}}]{S_{chi}}^{CL}(\omega)
\end{equation}

  We use the Metropolis method$^{5,6}$ for MC. We adopt the sequential updating procedure for MC updating in our calculation. As our model is highly frustrated, it takes long time to equilibriate. Therefore we discard all the spin configurations with MC age below $5000$ MC steps/spin (MCS/spin) for static calculations involving thermodynamic averages. We store spin configurations with MC ages between $8000-10,000$ MCS/spin, $14,000-16,000$ MCS/spin and $40,000-42,000$ MCS/spin for various calculations and investigations, depending upon the temperatures of interest and the model parameters. In particular these configurations are also used as boundary conditions for MD calculations. The configurations used for static calculations numbered $1000$ and those for MD numbered $5$. We use the lattice of size $9X9X9$ containing $729$ spins. The magnetic ordering temperature $(T_N)$ was determined by locating the peak for the specific heat $C_{v}(T)$ calculated by MC. We had first carried out our MC calculations with parameters corresponding to those of Kawamura$^{1}$ and found a very good agreement between our results and his. This was done as a reliability check on our calculations. 
The equation of motion in our model in the semiclassical approximation takes the form :-
\begin{eqnarray}
\frac{d{\bf s}_{i}}{dt}={t_{0}}^{-1}[\sum_{j(nn)}{\bf s}_{i}X{\bf s}_{j}+\frac{J_2}{J_1}
\sum_{j(nnn)}{\bf s}_{i}X{\bf s}_{j}+\frac{D}{J_1}[{-\bf x}{{\bf s}^{y}}_{i}+{\bf y}{{\bf s}_{i}}^{x}]] 
\end{eqnarray}
where $t_{0}$ is the natural time unit in our problem, where
\begin{equation}
{{t_{0}}^{-1}}={\frac{2\pi}{h}}J_{1}(S(S+1))^{\frac{1}{2}}
\end{equation}
The vector $\bf s_{i}$ represents a unit spin vector at the lattice site i with the operator ${\bf S}_{i}^{op}$ replaced by the classical vector ${\bf s}_{i}(S(S+1))^{\frac{1}{2}}$.
We calculate $S_{chi}(\omega)$ as a function of $\omega$ for our model at different temperatures both in the ordered phase as well as in the disordered phase. Later we also carried out MCMD calculation in the presence of a small magnetic field of strength $5$ Tesla applied along $x$ direction to verify the intrinsicity of this chiral mode.

  Regarding the study on trimerization, we take the following model,
\begin{equation}
J_{1}^{\prime}=J_{1}(1+(-1)^{\delta}A)
\end{equation}
 where $\delta$ takes the value even for the plaquette involving the points $[(x,y),(x+1,y+1),(x,y+1)]$ and the value odd otherwise; $A$ is a positive quantity denoting the enhancement/suppression factor of the nearest neighbor(nn) coupling and $J_{1}^{prime}$ is the modified nn coupling in the presence of trimerization. Moreover there are some special sets of values for $x$ and $y$ where this trimerization occurs$^{4}$. We adopt the same pattern in our calculation.
  All the simulation work reported in this paper was carried out using two machines. The MC calculations were done on Dec Alpha (IISC) and the MD calculations were carried out on a networked Linux machine  at Bose Centre. The total duration of each dynamical run in MD denoted by $t_{max}$  was chosen as $75t_0$ to get resolution width of $0.8$ mev.

\section{Results}

  In this section we present all our important results, for both the static and the dynamic properties. We have done calculations with two sets of model parameters which have same values for $J_{1}$ and $J_{2}$ but differ in the values of $D$. We calculated $T_{N}$ for both this model by MC calculation. The two sets of model parameters used by us are:-
\begin{equation}
 J_{1}=57.0K,\; J_{2}=5.7K,\; D=-5.7K
\end{equation}
 for the first system and
\begin{equation}
 J_{1}=57.0K,\; J_{2}=5.7K,\; D=-8.55K
\end{equation}
 for the second system. From our MC calculations we determine $T_N$ for both the systems. They are found to be:-
\begin{equation}
 T^{1}_{N}=66K,\;T^{2}_{N}=68K
\end{equation}
For trimerization study we chose the system parameters as
\begin{equation}
JN=1.0,\; JNN=1.0,\; D=0.0
\end{equation}
This gives 
\begin{equation}
T_{N}=1.2
\end{equation}
for the untrimerized case. For the trimerized case we choose $A$ to be equal to $0.2$. This gives the value of $T_{N}$ to be close to $1.0$.

 The sublattice magnetization vectors for all the $9$ layers in the ordered phase are determined from our MC calculations at the temperature of $20$K. These calculations have been done for $2$ MC samples prepared independently with MC ages varying between $14,000$ and $16,000$ MCS/spin. We have done this study only for the first system. The results for total magnetization vectors (taking into account the all 3 sublattice magnetization vectors) for sample $1$ and sample $2$  are presented in tables $1$ and $2$ respectively. It should be pointed out that the true values of the magnitudes of the magnetization can be obtained by dividing the corresponding magnitudes occurring in the tables 1 and 2 by a factor of 3. We have also studied the thermal  evolution of $\gamma$ and it is presented in table $3$. Moreover the effect of trimerization on $\gamma$ has also been studied for our model with $D$ put to zero. This result is displayed in table $4$. 

 Regarding the dynamical properties, we present both $S_{chi}(\omega)$ vs. $\omega$ and $S({\bf q}, {\omega})$ vs. $\omega$ for constant q-scans, obtained from our MCMD calculations, in the form of various figures. We consider both individual layers as well as layer averaged value in these graphs. We have carried out these calculations for our model with parameters appropriate to both first system and second system. The $\bf q$ vectors have been taken along both $<110>$ as well as $<111>$ direction for our calculation of $S(\bf q, \omega)$.

\section{Discussion}
\begin{table}
\caption{Total magnetization vectors for 9 layers for MC sample 1 corresponding to system 1 at 20K; {\bf M} denotes the total magnetization, {\bf$M_{xy}$} the total magnetization vector in the $xy$ plane etc., $\phi_{xy}$ denotes the angle of the vector {\bf$M_{xy}$} with the x-axis, $V_m$ represents the 3-d volume (with the sign) made by the 3 components of $\bf M$ and the angle $\phi_{xy}$ has been given in units of $\pi$.}
\begin{center}
\begin{tabular}{|c|l|c|c|c|c|}
\hline\hline
 LAYER INDEX &   $|M_{xy}|$ &    $\phi_{xy}$&      $M_z$ &      $V_m$ &   $|M|$ \\ \hline

  1          &    0.3     &    0.60    & -1.7        & 0.1     & $\sqrt{3.0}$ \\ \hline

  2         &     1.0     &    0.53    & 1.5    &  -0.2  &   $\sqrt{3.0}$  \\ \hline

  3        &    $\sqrt{2.0}$ &    0.44   &   -1.0  &   -0.4 &   $\sqrt{3.0}$ \\ \hline

  4        &      0.6     &    0.60  &   0.3   &   -0.04 &      $\sqrt{0.5}$\\
\hline

  5       &     $\sqrt{2.9}$ &     0.52 &   0.3   &   -0.05 &  $\sqrt{3.0}$\\
\hline

  6       &     $\sqrt{2.0}$ &     0.52 &   -0.8  &    0.1  &  $\sqrt{3.0}$\\
\hline

  7     &       1.1    &       0.53  &   1.1   &   -0.1 &  $\sqrt{3.0}$\\
\hline

  8     &        0.7   &        0.46  &     -1.5  &   -0.1 &  $\sqrt{3.0}$\\
\hline

  9     &       0.4    &       0.60   &    1.7   &  -0.07 &   $\sqrt{3.0}$\\
\hline
\end{tabular}
\end{center}
\end{table}

\begin{table}
\caption{Same as table 1 but for MC  sample 2 }
\begin{center}
\begin{tabular}{|c|l|c|c|c|c|}
\hline\hline
 LAYER INDEX &   $|M_{xy}|$  &  $\phi_{xy}$&  $M_z$ &  $V_m$  &  $|M|$ \\
\hline
  1          & 0.8           &  0.46      & 1.5    &   0.1   &  $\sqrt{3.0}$\\ \hline

  2          & $\sqrt{2.0}$ &    0.46    &   -1.0  &    -0.3 & $\sqrt{3.0}$\\ \hline

  3         &  $\sqrt{2.9}$ &    0.40    &    0.5  &     0.5 & $\sqrt{3.0}$\\ \hline

  4         &  $\sqrt{2.8}$ &    0.35    &    0.2  &     -0.3 &  $\sqrt{3.0}$\\ \hline

  5         &  $\sqrt{2.5}$ &    0.31    &  -0.7   &   -0.8 &  $\sqrt{3.0}$\\ \hline

  6         &  $\sqrt{1.6}$ &   0.35    &  1.2     &   0.8 &  $\sqrt{3.0}$\\ \hline

  7         &  $\sqrt{0.7}$ &   0.25    &  -1.5    &   -0.5 &  $\sqrt{3.0}$\\ \hline

  8         &  0.1 &         0.25   &   1.6    &     0.02 &  $\sqrt{2.6}$\\ \hline

  9        &   0.3  &        0.40  &    -1.5   &    -0.05 &  $\sqrt{2.0}$\\ \hline

\hline
\end{tabular}
\end{center}
\end{table}

\begin{table}
\caption{$\gamma(T)$ for system 1 in the untrimerized case}
\begin{center}
\begin{tabular}{|c|c|}
\hline\hline
    Temperature(T)   &                   $\gamma$\\  \hline
     (in K)          &                          \\ \hline
       67            &                      0.04\\ \hline
       66            &                      0.20\\ \hline
       65            &                      0.50\\ \hline
       64            &                      0.20\\ \hline
       62            &                      0.04\\ \hline
       60            &                      0.14\\ \hline
\hline
\end{tabular}
\end{center}
\end{table}

\begin{table}
\caption{$\gamma(T)$ for system 1 in the trimerized case}
\begin{center}
\begin{tabular}{|c|c|c|}
\hline\hline
      Temperature(T)     &               $\gamma$ (trimerized) &    $\gamma$(untrim.) \\ \hline 

          0.5            &                    0.30          &      0.20\\ \hline     
          0.1            &                    0.14          &\\         \\ \hline
\hline
\end{tabular}
\end{center}
\end{table}
\begin{table}
\caption{$<{\omega}_p>_{layav}(T)$ vs. $T$ for both system 1 and system 2}
\begin{center}
\begin{tabular}{|c|c|c|}
\hline\hline
        $D(k)$         &          $T/T_N$       & $<{\omega}_p>_{av}(T)$ ($mev/hcross$)\\ \hline
         -5.7          &          0.30          &        0.40\\ \hline
         -5.7          &          0.50          &        0.13\\ \hline
         -5.7          &          0.75          &        0.13\\ \hline
         -5.7          &          1.10          &        0.04\\ \hline
         \hline
         -8.55         &          0.25          &        0.50\\ \hline
         -8.55         &          0.75          &        0.40\\ \hline
         -8.55         &          1.10          &        0.50\\ \hline
\hline
\end{tabular}
\end{center}
\end{table}

\begin{figure}
\centering
\psfig{file=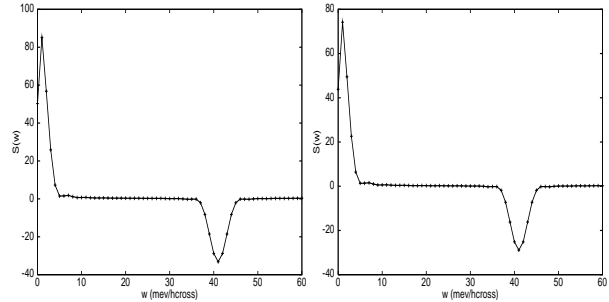,height=4cm,angle=0}
\caption{$S_{chi}(\omega)$ vs. $\omega$  corresponding  to system 1 at temperature of $0.3T_N$; for layer 1 (left graph) and for averaging over all layers (right graph)}
\end{figure}

\begin{figure}
\centering
\psfig{file=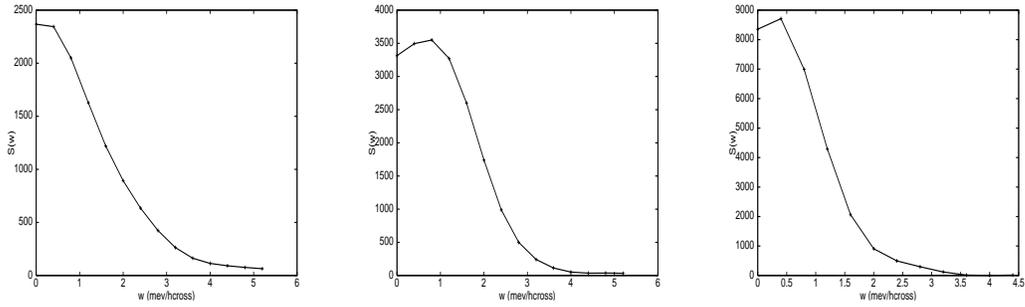,height=4cm,angle=0}
\caption{$S_{chi}(\omega)$ vs. $\omega$ corresponding to system 2; 
for layer 1 at temperature of $0.75T_N$ (left most graph), 
for layer 3 at temperature of $0.75T_N$ (centre graph) and 
for any layer at temperature of $0.25T_N$ (right most graph)} 
\end{figure}

\begin{figure}
\centering
\psfig{file=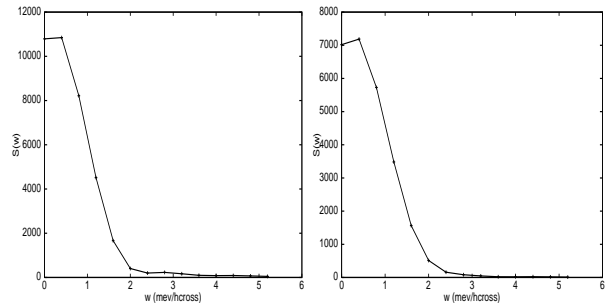,height=4cm,angle=0}
\caption{$S_{chi}(\omega)$ vs. $\omega$ corresponding to system 1; for layer 1 in the presence of magnetic field of 5 Tesla at temperature of $0.5T_N$ (left graph) and for any layer in the absence of any magnetic field at temperature of $0.3T_N$ (right graph)}
\end{figure}

\begin{figure}
\centering
\psfig{file=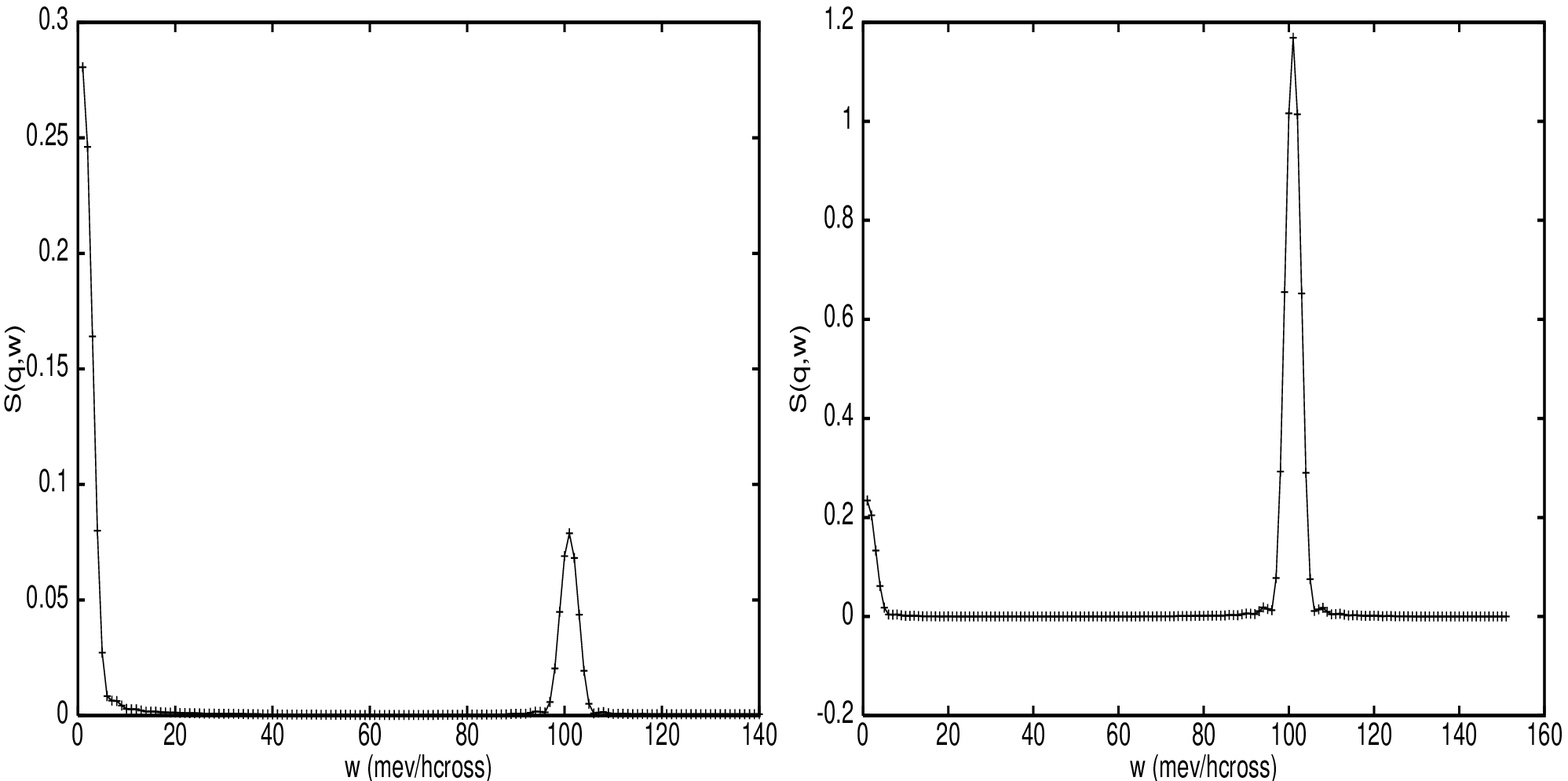,height=4cm,angle=0}
\caption{$S(\bf q, \omega)$ vs. $\omega$ corresponding to system 1 for layer 1 at temperature of $0.3T_N$; the values of $\bf q$ are $(0,0,0)$ (left graph) and $(6,6,0)$ (right graph)}
\end{figure}

 From tables $1$ and $2$ we see that there is a non-trivial correlations amongst $\bf M$ occurring in the 9 layers. Firstly $M_{z}$ shows anti-parallel correlations between adjacent layers, excepting two layers, for both the MC samples. The same thing cannot however be said regarding $M_{xy}$. In sample $2$ it mostly shows an antiparallel arrangement but not so for sample $1$. Moreover in both sample $1$ as well as in sample $2$, a few layers exhibit a large suppression of total magnetic moment. The value of $\gamma$ also shows variation with layer index but indicates a "fairly high degree of coplanarity"  for the strictly non-coplanar $3$-sublattice structure of magnetic ordering, for all the layers for both the MC samples. This result is of much significance in the presence of uniaxial anisotropy. In contrast to the case with pure Heisenberg model however, the angles between the $3$ sublattice ordering directions are not $120$ degrees in general in our case.  From the results presented in table $3$ we see that the degree of coplanarity (layer averaged) for system $1$, has an oscillatory behavior as the temperature is lowered from $T_{N}$. From table $4$ we see that trimerization causes coplanarity to decrease a bit in the absence of the anisotropic term. Moreover we also observed that $3$-sublattice structure is much weaker i.e. much less distinct in the case of trimerized model compared to the untrimerized case. These results support the conjectures of the experimentalists$^{4}$. We also see that with decrease in temperature the degree of coplanarity increases even for the trimerized system. This study also brought out the fact that trimerization causes $T_{N}$ to drop.

  Regarding spin dynamics, we see many more interesting features in $S_{chi}(\omega)$ than in $S(\bf q, \omega)$. There exists a well defined propagating mode for $S_{chi}(\omega)$ in the ordered phase for both the sets of system parameters, as are found from the figures 1 and 2. These graphs describe the behavior of $S_{chi}(\omega)$ as a function of $\omega$ for some particular layers. The frequencies of propgating modes i.e. the values of the frequencies at which the curves show the peaks, are found to be dependent on the layers, for both the systems. The layer averaged peak frequencies have also been calculated. These frequencies denoted by $<{\omega}_{p}>$ have been studied as a function of temperature for both the systems. The results of this study are displayed in table $5$. These modes disappear in the disordered phase i.e. for $T$ above $T_N$, excepting in the case of system $2$. It should be pointed out that these $<{\omega}_{p}>$'s are in general different from the peak position of the layer averaged $S_{chi}(\omega)$. Another notable feature in these figures is the existence of a dip at higher value of $\omega$ and that occurs in the regime of -ve values for $S_{chi}$. This regime probably describes the spin flipping process causing the chirality to be reversed at high energies.
  From the analysis of these results we see a very clear thermal evolution of the chiral mode for system 1. The softening of averaged peak frequency with increase in temperature is seen quite distinctly for system1. For system $2$ with higher value for the magnitude of $D$ however, we hardly see any thermal evolution for the peak frequency. This is probably due to the freezing of spin dynamics by the rather large value of $D$ in the low $\omega$ region. It is to be noted that our $S_{chi}(\omega)$ is actually $S_{chi}(q=0,\omega)$. The intrinsicity i.e. non-spurious nature of these modes were verified by applying a magnetic field, as has been described earlier. The peaks were found to essentially vanish and become "shoulders"  in the presence of applied field, in contrast  to the zero field situation. This is brought out in figure $3$. The existence of these chiral modes support the theoretical conjecture$^{4}$. At low temperature the layer averaged peak frequency corresponding to this chiral mode, seems to scale as $D^{0.5}$, with the same values for $J$'s and $\frac{T}{T_{N}}$. It is important to add that these chiral modes disappear in the disordered phase i.e. for $T$ above $T_N$, as we observed in our MCMD calculation for system $1$ at the temperature of $1.1T_N$. We however have not presented the results here.

  The total spin dynamics described by the full dynamic structure function $S(\bf q, \omega)$ as a function of $\omega$ in the constant-$\bf q$ scan, exhibit spin wave like collective modes in the ordered phase for both the systems. These modes however look quite dispersionless, both with respect to the magnitude and direction of $\bf q$. They do not show any appreciable temperature dependence (softening) in the ordered phase either. Besides this finite frequency peak, we also see a central peak i.e. a peak at zero frequency. The relative intensities at these two peaks are however a function of temperature and the value of $q$. These results are presented in the figure $4$. The finite frequency peak is quite sharp at low temperature and is of much higher intensity than the central peak for large values of $|q|$, as we can see from these figures. The central peak arises probably from the very slow dynamics of the longitudinal components of the spin, the longitudinal directions being the $3$-sublattice ordering directions. For low values of $|q|$ however, the central peak dominates over the finite $\omega$ peak in terms of intensity. The apparent dispersionless and gapped behavior of these "spin waves" are probably due to the presence of $D$ term. The observed insensitivity of these modes to temperature in the ordered phase also arises from this anisotropy term, most likely.

  It is quite striking to observe the contrast in the behavior of $S_{chi}(\omega)$ and $S(\bf q, \omega)$ particularly with respect to thermal dependence. This probably reflects the lesser influence the single ion anisotropy has on $2$-plaquette correlations than on $2$-spin correlations. Moreover no central peak occurs in the presence of the finite frequency peak for $S_{chi}(\omega)$ in contrast to the case with $S(\bf q, \omega)$. It is to be noted that the position of the "spin wave" peak looks consistent with the magnitude of $J_{1}$. It is extremely interesting to notice that the position of the finite frequency peak for $S(\bf q, \omega)$ is very close to the position of the dip for $S_{chi}(\omega)$. On the other hand the peak position of $S_{chi}(\omega)$ is close to the positon of the central peak for $S(\bf q, \omega)$. This probably implies that the dominant contribution to the total spin dynamics here comes from the chiral dynamics. It would also be interesting to study the dynamic structure factor corresponding to the staggered chiral correlator $S_{chi}(\bf q, \omega)$ by MCMD. In brief the spin dynamics of a geometrically frustrated system in the presence of single ion anisotropy is very rich. It would be quite exciting to determine the role and importance of quantum fluctuations in these systems ($S=2$) by comparing our semiclassical results for dynamic structure factors with those from inelastic neutron scattering experiments in a detailed quantitative way . Unfortunately detailed experimental results for these systems are still awaited.

\section{Acknowledgments}

  This investigation was initiated when the author was visiting Indian Institute of Science (IISC), Bangalore. He would like to thank B.S. Shastry for drawing his attention to this problem and for his constant encouragement during the course of this investigation. He would also like to thank the department of Physics, IISC for the kind hospitality and  Sreenivasan for very kind help with the computer facilities for part of the work carried out. The author would also like to acknowledge the assistance of Durga Paudyal in the preparation of the manuscript. 

\section{References}

\noindent 1  \\
\vspace{0.5cm}
 H.Kawamura, J.Phys.Soc. Japan 58, 584 (1989).  \\
\noindent 2  \\
\vspace{0.5cm}
 D.V. Spirin, J. Mag. and Magnetic Mat. 264, 121 (2003).  \\
\noindent 3  \\
\vspace{0.5cm}
 M.F. Collins  and O. Petrenko, Can. J. Phys., 75, 605 (1997); V. P. Plakhty et al, Europhys Lett. 48(2), 215 (1999).  \\
\noindent 4  \\
\vspace{0.5cm}
 B.S.Shastry, Private Communications, (2002); S.V. Maleyev et al, J. Phys.: Condensed Matter, 10, 951 (1998).\\
\noindent 5  \\
\vspace{0.5cm}
 R.Chaudhury and B.S. Shastry, Phys. Rev. B 37, 5216 (1988).  \\
\noindent 6  \\
\vspace{0.5cm}
 Ranjan Chaudhury and B.S. Shastry, Phys. Rev. B 40, 5036 (1989).  \\

\end{document}